# Increasing Security in Cloud Environment

Priyanka Naik[1*], Sugata Sanyal[2]

[1]Computer Science Department, Manipal Institute of Technology, Manipal, India

ppnaik1890@gmail.com

[2] Corporate Technology Office, Tata Consultancy Services, Mumbai, India

sugata.sanyal@tcs.com

*Corresponding Author

**Abstract**

The concept of cloud computing was introduced to meet the increase in demand for new application for a project, and to provide a large storage facility whenever or wherever a user needs it. The cloud system facility helped many industries as well as individual users to get authentic software at a very low cost. But with this new system comes the major concern of security, as the connection to the cloud is through the web and the data and application availability need to be handled for each client. The paper describes the various security measures that can be added in isolation or in combination for securing data transmission, server and client.

## 1. Introduction

The concept of thin client in cloud computing has revolutionized its uses as it provided individuals as well as company users an efficient cloud computing [1]. Security is a major concern for the cloud environment. Incorporating confidentiality, integrity, and availability features in the cloud system is necessary. Protecting user identity, user data, industry data privacy is needed for a secure cloud computing. Data theft can happen during transmission as well as when stored on the server. So, transmission also needs to be protected. Pal et al. proposed the idea of using a virtual machine monitoring system to monitor the cloud system [2]. Malicious users need to be prevented from exploiting the cloud system. This paper presents various ways that can be incorporated for securing and protecting the cloud computing.

This paper has four sections in addition to introduction and conclusion section. Section 2 covers approaches for transmission security, while section 3 and 4 deal with server and client security respectively. Section 5 covers advantages of various security measures.

## 2. Securing the Transmission

The data communication between the client and cloud server passes through the network, where it can be exploited. Malicious flooding routes need to be handled by setting a limit for each route [3]. The number of packets a route can transfer defines its limit. To secure the transmission signals encryption algorithms i.e. public key and private key encryption, can be used with spread spectrum modulation [4]. Wireless transmission can be secured using Wired Equivalent privacy (WEP), SSID for each access point and MAC address filtering [5].

### *2.1 Tunneling*

The concept of tunneling can be used to secure the data during transmission. The packet destined for the cloud server can be encapsulated in a packet with the address of a different node. Packets on reaching this node will be redirected to the server by the node. This encapsulation prevents the attacker to track down packets meant for the server thus reducing their possibility of getting hacked. The threat of goggle hacking can also be reduced by incorporating this methodology [6].

### *2.2 Use of Virtual Circuits*

The IP packets are transferred as datagrams through the network, so they follow the best path possible. However, they may pass through the router which may have been attacked by an attacker. In such cases, the packets can be traced down and exploited by the attacker. To prevent this attack a virtual circuit can be implemented. In this methodology, the server during the connection establishment sets a fixed route, which the data packets need to follow. This path is through authorized routers. This path insures data security but may fail if a router on the path is down.

## 3. Securing the Servers

### *3.1 Intrusion detection system (IDS)*

With the use of intrusion detection system attacks like SQL injection can be monitored. The IDS can keep track of possible user requests and queries to monitor these attacks [7]. The IDS can be considered as an immune system for the system. Combining it with soft computing techniques it can detect intrusions in the network [8]. Dimensions such as source and destination IP addresses, port addresses, CPU cycles etc can be used to detect an ambiguous behavior in the network [9]. Replication of state machine can help to detect the nodes in the state with abnormalities, which indicate the attacked nodes [10]. For mobile ad-hoc networks, the dynamic source routing based algorithms can be employed to detect a possible intrusion [11]. The concept of fuzzy networks can be used to analyze the network [12].

*3.2 Separate Servers*

Clubbing multiple applications on a single server increases the load on a server, and also makes it more vulnerable to different attacks [13]. The running of one application can pose a threat to other applications on that server. Solution to this is storing different applications on different servers. This prevents conflicts but increases the cost of setting up the servers. Multi threading parallelization can be used to speed up the servers thus reducing the response time [14]. With the introduction of IPV6, address assignment and resolution problem has reduced [15].

*3.3 Store Hashed Values*

The data stored in the cloud databases is mostly plaintext. If the database is broken then the entire data can be exploited. Storing the hashed value of data can prevent this exploitation. The key for hashing can be securely stored to prevent the exploitation. The key can be randomly generated for each database object like table and view. An alternative way is to use the sum of prime numbers or natural numbers to embed the messages without much distortion [16]. The servers can be made computationally strong by using grid computing. Grid security can then be added to enhance the security of the server [17].

*3.4 Replication*

To make the data available to the user whenever he/she needs it makes it necessary for the server to be working without fail. But in actual scenario it is very difficult for the server to be working all the time. Some failure can lead to the server to be down leading to its unavailability to service. Replication helps to handle the failure. When one server is not working the replicated server can be used. Synchronization is the major concern of replication.  Periodic updation can help to keep all the servers up to date.

*3.5 Threshold for Server Load*

The main purpose of denial of service attack is to overload the server so that it crashes and thus preventing the legitimate user from accessing the server. The solution to this is setting a threshold value for the load a server can handle [18]. When the number of requests serviced by the server reaches threshold the rest of the load is transferred to the replicated servers. The server load can be managed effectively in this way. Locking of data in case of write to the same file, from request to both the servers helps to maintain the integrity. Flooding request also need to be handled to prevent the server from getting overloaded.

## 4. Securing the Client

Client data can be secured by combining attribute based encryption and proxy encryption. Client's access accounting should be done [19].

### *4.1 Digital Signatures*

Now a days digital signature is used to authenticate the server systems. But no such authentication is available on the client side. In the cloud environment the clients are known beforehand, so identifying them using digital signatures is possible. Different methods for signature like RSA can be used. This increases the computational time but helps to maintain the security. The large storage of the cloud system can effectively help in RSA computation, thus preventing the client machine from being overloaded [20]. Digital water marking can also be used as a method for unique identification [21].

### *4.2 One Time Password*

Ones an attacker gets access to the password of a user he can use replay attack to exploit the system. One time password help prevent the replay attacks. One time passwords are randomly generated passwords by the server sent to the client through a secure channel that is used by the client. Client gets a unique password for each session. Embedding the message by exploiting the redundancies in html pages is a possible solution if one time password is proving expensive [22]. One time passwords also secure client's data from cookie poisoning as passwords change with each session and so the information stored in the system cannot be reused [12].

### *4.3 Authenticate for each write*

Each write means permanent storage to the database. So, it should be from a valid client. Asking for a password before each storage request helps to ensure that the changes to the file are made by authentic users only. Multifactor authentication can be used to increase the security [23]. Hashing the password method can be deployed for authenticating the users [24].

*4.4 Distributed Storage*

Distributed storage means storing parts of the client's data on different location. Even if one location is attacked the client would not loose the entire data. Hence, restoration will become easier. The client need not know at which location his/her data is stored. The system can be location transparent. Making the system distributed add up the issues related to distributed storage. But they can be handled with proper synchronization. To secure the data storage, erasure correcting code in file distribution should be used [25].

*4.5 Local Servers*

Frequent users use some common applications. They mostly need only few applications. Some local server can be used which store the frequently used applications. This helps to avoid the network congestion and therefore get fast access. Local servers can also cache the data for sessions and template for dynamic documents [26].

*4.6 Temporary storing on local disk*

The threat of attack can be minimized by avoiding a constant connection to the internet. This can be achieved by temporarily storing the file on local disk for one session and updating the cloud database once all the operations for that session are done. Thus there is no need for a constant connection to the internet thus reducing the threat of an attack. This requires the local disk to be large enough to store the session's information. The disconnection from the internet during the execution time reduces the threat of bot attack. Methods to counter this attack can be used when the client is communicating to the server via the internet services [27].

**5. Building the Secure Cloud**

The cloud environment can be secured by combining the features mentioned above depending on the level of security needed. Two factor authentications can be done by using one time password as well as digital signature. Advantages of such combination are given in Table 1.

**Table 1: Advantages of the Security Measures for cloud environment**

| Security Measures | Advantages |
|---|---|
| Two factor authentication with digital signature and one time password | Secures system without the need of any extra hardware. |
| Replicating storage of hashed data | Secured data available during failure. |
| Authenticating for each write with digital signature | Unique identification of client. Secured write. |
| Use of virtual circuit along with IDS | Protect an attack during transmission and detect any attack in the code which reaches the server. |

## 6. Conclusion

Cloud computing is used to aid users by providing software, storage as services. But with this high availability comes the threat of being attacked. This system is vulnerable to attack as all the tasks are performed over the internet, which is highly insecure. The security of the system can be improved by including additional features in the transmission medium, the server side and the client side, keeping in view the security aspects. Transmission channel can be secured by tunneling. The server side security can be enhanced by using appropriate IDS, hashing or replication. The client can be secured by proper authentication technique like digital signature. Cloud environment is a boon to the IT industry as well as individual users but if the security measures are inadequate then the entire system's concept will fail. Therefore all possible ways to increase the security should be taken into consideration.